\documentclass[12pt]{article}
\usepackage{mathtools}
    \allowdisplaybreaks
\usepackage{amssymb}
\usepackage{bm}
\usepackage{physics}
\usepackage[mathscr]{euscript}
\usepackage{multirow}
\usepackage{enumitem}
\usepackage{xcolor}
\usepackage[colorlinks,citecolor=blue,linkcolor=blue,urlcolor=blue,pagebackref]{hyperref}
\usepackage{amsthm}
    \theoremstyle{plain} %

        \newtheorem{lemma}{Lemma}
        \newtheorem{proposition}{Proposition}
    \theoremstyle{definition}
        \newtheorem{assumption}{Assumption}
        \newtheorem{definition}{Definition}
        
    \theoremstyle{remark}
        
        \newtheorem{remark}{Remark}
\usepackage[margin=1in]{geometry}
\usepackage{booktabs}
\usepackage{microtype}
\usepackage{setspace}
\usepackage[ruled]{algorithm2e}
\usepackage[labelfont=bf, font=small]{caption}
\usepackage{natbib}
    \bibpunct[, ]{(}{)}{,}{a}{}{,}%
\usepackage{tikz}

\begin{document}

\begin{titlepage}
\title{Seesaw Experimentation: A/B Tests with Spillovers \thanks{Contact information: Jin Li (\url{jli1@hku.hk}), Ye Luo (\url{kurtluo@hku.hk}), Faculty of Business and Economics, The University of Hong Kong, Pokfulam Road, Hong Kong SAR.
Xiaowei Zhang (\url{xiaoweiz@ust.hk}),
Department of Industrial Engineering and Decision Analytics, The Hong Kong University of Science and Technology, Clear Water Bay, Hong Kong SAR. We thank Shan Huang, Hanzhe Li, and Ruohan Zhan for helpful comments and discussions.} }
\author{Jin Li \and Ye Luo \and Xiaowei Zhang}
\date{}
\maketitle
\begin{abstract}
This paper examines how spillover effects in A/B testing can impede organizational progress and develops strategies for mitigating these challenges. We identify a phenomenon termed \emph{seesaw experimentation}, where a firm's overall performance paradoxically deteriorates despite achieving continuous improvements in measured A/B testing metrics. Seesaw experimentation arises when successful innovations in primary metrics generate negative externalities in secondary, unmeasured dimensions. To address this problem, we propose implementing a positive hurdle rate for A/B test approval. We derive the optimal hurdle rate, offering a simple solution that preserves decentralized experimentation while mitigating negative spillovers.

\vspace{15mm}


\end{abstract}
\setcounter{page}{0}
\thispagestyle{empty}
\end{titlepage}
\pagebreak \newpage

\section{Introduction}

A/B testing has become an important tool for modern business, enabling firms to drive innovation, improve user experiences, and increase revenue growth \citep{Kohavi20Trustworthy,KoningHasanChatterji22}. Major tech firms such as Amazon, Google, Meta, and Microsoft deploy thousands of such tests annually, embedding data-driven decision-making into their innovation processes \citep{Thomke20Experimentation,azevedo2019abtesting}.

The massive scale of experimentation in modern organizations requires A/B testing to be initiated with a decentralized structure. While central analytics teams execute these experiments, decisions about what to test and which metrics to track are distributed across product teams, business units, and regional divisions. This decentralization enables rapid innovation but complicates the measurement and management of cross-team effects.

A key challenge of decentralized testing is the occurrence of \emph{spillover effects}—unintended consequences that arise when teams' successful A/B test innovations inadvertently harm other parts of the business. These spillovers manifest in various ways: improvements in one product can decrease engagement with others as users redistribute their attention, online sales growth might cannibalize offline revenue, and improving advertising revenue can worsen user experience. Such effects are particularly pronounced on large digital platforms, where products and services are tightly integrated within the same ecosystem. Yet despite their prevalence, these spillover effects remain difficult to measure and often go untracked in the testing process\footnote{Industry experts at a leading Chinese digital platform confirm that cross-product externalities are rarely tracked systematically in A/B testing.}.

The purpose of this paper is to study A/B testing in the presence of spillover effects. We explore how these spillovers\footnote{We mostly focus on negative spillover effects in this paper.} can fundamentally slow organizational progress and what firms can do to mitigate them. Our first contribution is to highlight and identify \emph{seesaw experimentation}: a phenomenon in which the overall performance of the firm deteriorates over time even as it achieves continuous improvements in the measured metrics of A/B testing.

The name ``seesaw" captures the oscillating nature of this phenomenon. Just as one end of a seesaw rises while the other falls, advances in one dimension often come at the cost of decline in others. When firms implement innovations to boost their primary metrics—whether in specific products, channels, or KPIs—these ``improvements" can inflict hidden damage across other areas of the business. As strategic priorities shift, from gross merchandise value today to user satisfaction tomorrow, new innovations may actually reverse previous gains. This creates a dangerous illusion of progress: while the firm is marching from victory to victory, its overall performance quietly declines.

To formally analyze this phenomenon, we develop a bivariate normal distribution framework (later extended to distributions with fat tails) that demonstrates the emergence of seesaw experimentation. Our model identifies specific conditions where firms achieve consistent A/B test successes yet face declining overall performance. Two key factors drive this phenomenon. First, a higher signal-to-noise ratio (the absolute value of mean over standard deviation) increases the likelihood of seesaw experimentation. Second, this effect strengthens as the correlation between dimensions becomes more negative.

Our second contribution is to develop a practical solution for managing spillover effects through a positive hurdle rate for A/B test approval. Rather than implementing all positive innovations, this approach only adopts changes that exceed a specified performance threshold. We analytically derive the optimal hurdle rate that balances gains in the primary dimension against losses in the secondary dimension. The optimal threshold emerges at the point where marginal benefits equal the expected marginal costs in the secondary dimension.

Our framework provides actionable guidance by characterizing how the optimal hurdle rate varies with key parameters of the performance distribution, including the correlation between primary and secondary effects and the means and variance of the distributions. Relative to other potential solutions to deal with the spillover effect, our solution maintains the autonomy of decentralized A/B testing by avoiding the need for complex cross-team coordination mechanisms or comprehensive measurement systems.

A/B testing and experimental design in the digital economy has drawn significant interest from academia and industry; see \cite{Zhao24} for a recent survey.
However, the literature addressing our specific focus is relatively small. \cite{berman2018phacking} examine how p-hacking in A/B testing can lead to false discoveries and impair experimental improvements. \cite{azevedo2019abtesting} analyze how distributional characteristics, particularly fat tails, influence experimentation allocation strategies.\footnote{We extend our analysis to fat-tailed distributions in the Appendix.} \cite{McClellan2022} explores how agency problems affect A/B test adoption mechanisms. While these studies provide valuable insights, they primarily focus on single-dimensional outcomes. Our work departs from this literature by focusing on the interconnectedness of performance across multiple dimensions, emphasizing how improvements in one dimension can generate spillover effects in others.

Our work also connects to the broader literature on measurement error in economics and organizations. Classical work has long recognized how imperfect measurement can distort decision-making \citep{Heckman79,HolmstromMilgrom91}. These measurement challenges arise from both organizational capacity constraints \citep{Ocasio97} and incentive issues \citep{Kerr75}. Such challenges become particularly salient in the digital age, where algorithm decision-making can exacerbate the measurement issue \citep{LiLuoZhang22_fullversion,LiLuoZhang23_bandit}. In our paper, measurement problems at the organizational level can lead the firm to a false sense of progress, even if its performance declines over time.

\section{Model Formulation}

Consider a firm whose management is concerned with two performance dimensions, denoted as $u$ and $v$. In each period $t = 1,\ldots,T$, the firm identifies its strategic priority $a_t \in \{u, v\}$ between these two dimensions.
We refer to this prioritized dimension as the \emph{primary} dimension, and the other as the \emph{secondary} dimension.

The firm then conducts an A/B test to evaluate a potential innovation using a metric aligned with the primary dimension $a_t$, measuring the innovation's effect in this dimension.
Let $u_t$ and $v_t$ represent the innovation's effects on dimensions $u$ and $v$, respectively.  Note that only the effect on the primary dimension is measured in the A/B test.

Subsequently, the firm decides whether to adopt this innovation based on the measured effect.
Let $d_t \in \{0,1\}$ denote the adoption decision in period $t$, where $1$ indicates adoption and $0$ indicates non-adoption.
The decision rule can be expressed as:
\[d_t = \mathbb{I}(a_t = u, u_t>0) + \mathbb{I}(a_t = v, v_t >0).\]
In other words, the firm adopts the innovation if and only if it improves the firm's performance in the primary dimension $a_t$ for the current period $t$.

In this paper, we examine the firm's \emph{cumulative performance} through a multi-period innovation process driven by A/B testing.
Let
\[U_T = \sum_{t=1}^T d_t u_t \quad\mbox{and}\quad V_T = \sum_{t=1}^T d_t v_t.\]
Here, $U_T$ represents the cumulative performance in dimension $u$ of the \emph{adopted innovations} over the time horizon $T$, while $V_T$ represents the same for dimension $v$. We define the firm's \emph{overall performance} at time $T$ as the sum of these cumulative performances: $U_T + V_T$.

Suppose that the market environment changes exogenously and management accurately identifies strategic priorities to adapt over time.
We model this by treating the strategic priorities $a_t$ as independent and identically distributed (i.i.d.).
Furthermore, suppose that each potential innovation's effects $(u_t, v_t)$ on both dimensions are i.i.d. across time periods, while allowing for correlation between dimensions within each period.

Formally, we make the following assumption throughout the paper:

\begin{assumption}\label{assump:iid}
\begin{enumerate}[label=(\roman*), noitemsep, topsep=0pt]
    \item $\{a_t:t \geq 1\}$ is a sequence of i.i.d. realizations of a random variable $A$, where $\mathbb{P}(A = u) = p_u$, $\mathbb{P}(A = v) = p_v$, and $p_u + p_v = 1$.
    $\{(u_t, v_t):t\geq 1\}$ is a sequence of i.i.d. realizations of a random vector $(U, V)$.
    \label{cond:iid}
    \item The sequences $\{a_t:t\geq 1\}$ and $\{(u_t, v_t):t \geq 1\}$ are independent of each other. \label{cond:indep}
\end{enumerate}

\end{assumption}

Under Assumption \ref{assump:iid},
the firm's overall performance is a sum of i.i.d. random variables:
\[U_T + V_T = \sum_{t=1}^T [d_t (u_t + v_t)].\]
The strong law of large numbers then implies that
\[
\mathbb{P}\left(\lim_{T\to\infty} \frac{U_T + V_T}{T} = \mathbb{E}[D (U + V)] \right) = 1.
\]

\begin{definition}
  A firm exhibits the phenomenon of \emph{seesaw experimentation} when its long-run average performance $\mathbb{E}[D (U + V)]$ is negative despite adopting innovations only when they demonstrate positive effects in their respective primary dimensions.
\end{definition}

For simplicity, we start by considering bivariate normal distributions. We extend our results to a wider class of distributions later.

\begin{assumption}\label{assump:normal}
\begin{enumerate}[label=(\roman*), noitemsep, topsep=0pt]

   \item $(U, V)$ follows a bivariate normal distribution with mean vector $m$ and covariance matrix $\Sigma$:
    \[m = \begin{pmatrix} \mu_u \\ \mu_v \end{pmatrix}\quad \mbox{and}\quad \Sigma =  \begin{pmatrix}\sigma_u^2 & \rho \sigma_u \sigma_v \\  \rho \sigma_u \sigma_v & \sigma_v^2\end{pmatrix}. \]
    Here, $\sigma_u^2$ and $\sigma_v^2$ are the marginal variances of $U$ and $V$, respectively, and $\rho$ is their correlation. \label{cond:normal}
    \item Both $\mu_u$ and $\mu_v$ are negative.
    \label{cond:neg-mean}
\end{enumerate}
\end{assumption}

The model uses several simplifying assumptions to highlight its core mechanism. We now examine how alternative assumptions affect the results.

\begin{remark}
    Our baseline model assumes one A/B test per period. This assumption can be generalized in several ways. The firm could conduct multiple tests simultaneously across both dimensions, or implement tests probabilistically based on historical outcomes. These extensions preserve our core findings.
\end{remark}

\begin{remark}
  The firm's adoption rule in our model depends entirely on the primary dimension outcome. When the firm could incorporate both dimensions--such as implementing changes when their weighted average is positive--this would weaken or eliminate the spillover effect. However, the spillover effect persists whenever the firm weighs the primary dimension more heavily, which is likely in practice given business pressures and test initiators' incentives to prioritize primary metrics.
\end{remark}

\begin{remark}
We have assumed a bivariate normal distribution, and as we will discuss later, our results extend to settings with multiple dimension and fat-tailed distributions. Within the bivariate normal distribution, Assumptions \ref{assump:iid} and \ref{assump:normal} can be further relaxed. For example, in Assumption \ref{assump:iid},  $a_t$ can be correlated over time. In addition, Assumption~\ref{assump:normal}\ref{cond:neg-mean} is not essential: seesaw experimentation can still occur when either $\mu_u$ or $\mu_v$ is non-negative while the other is negative, provided that $\mu_u + \mu_v < 0$.
\end{remark}

\section{Analysis} \label{section: symmetric}

We first consider the scenario where the two dimensions are symmetric.
In other words, while they can be correlated, $U$ and $V$ have the same marginal distribution:  $\mu_u = \mu_v$ and $\sigma_u = \sigma_v$.
We use $\mu$ and $\sigma$ to represent their common values, respectively.

Using the law of iterated expectations and the independence between $A$ and $(U, V)$, we can express $\mathbb{E}[D(U + V)]$ as:
\begin{align*}
& \mathbb{E}[D(U + V)] \\
={}& \underbrace{\mathbb{P}(A=u)}_{\text{Prob. of prioritizing dim. $u$}} \times \underbrace{\mathbb{P}(U>0)}_{\text{Prob. of adopting inno. in dim. $u$}} \times \underbrace{\mathbb{E}[U+V | U>0]}_{\text{Combined perf. if adopting inno. in dim. $u$}} \\
+{}& \underbrace{\mathbb{P}(A=v)}_{\text{Prob. of prioritizing dim. $v$}} \times \underbrace{\mathbb{P}(V>0)}_{\text{Prob. of adopting inno. in dim. $v$}} \times \underbrace{\mathbb{E}[U+V | V>0]}_{\text{Combined perf. if adopting inno. in dim. $v$}}.
\end{align*}
Hence,
a sufficient condition
for $\mathbb{E}[D(U + V)]$ to be negative is that both $\mathbb{E}[U+V | U>0]$ and $\mathbb{E}[U+V | V>0]$ are negative.
These quantities can be calculated explicitly under the bivariate normal assumption.

\begin{proposition}\label{prop:seesaw-sym}
Suppose Assumptions~\ref{assump:iid} and \ref{assump:normal} hold and $(U, V)$ follows a  symmetric distribution. If
\begin{equation}\label{eq:upper-bound-normal}
-1 \leq \rho < 2\left(\frac{-\mu}{\sigma}\right) M\left(\frac{-\mu}{\sigma} \right) - 1,
\end{equation}
then $\mathbb{E}[D(U + V)] <0$,
where $M(\alpha) \coloneqq (1 - \Phi(\alpha))/\phi(\alpha)$ is the \emph{Mills ratio} of the standard normal distribution,
with $\phi$ and $\Phi$ representing the probability density function and cumulative distribution function of the standard normal distribution, respectively.

\end{proposition}

Proposition~\ref{prop:seesaw-sym} shows that seesaw experimentation can occur. That is, while a firm may seemingly improve its performance constantly through successful A/B tests, its overall performance may deteriorate over time. This decline stems from significant negative spillover effects on secondary dimensions that offset the gain in the primary dimension.

Proposition~\ref{prop:seesaw-sym} also provides a sufficient condition for seesaw experimentation to occur. One factor is the signal-to-noise ratio $\alpha=|\mu|/\sigma$. (Note: $\mu$ is assumed to be negative.) We show in the Appendix (Lemma~\ref{lemma:Mills}) that $\alpha M(\alpha)$ is an increasing function of $\alpha$, implying that the sufficient condition is more likely to be satisfied when signal-to-noise ratio increases. This can happen in two ways: either through an increase in $|\mu|$ or a decrease in $\sigma$. The intuition behind this is straightforward. When $|\mu|$ is larger, the negative impact on the secondary dimension becomes more pronounced. Alternatively, when $\sigma$ is smaller, there's less variability in outcomes, which typically results in smaller average benefits in the primary dimension. Both scenarios make seesaw experimentation more likely to occur.

Another key factor is the correlation coefficient $\rho$ between the two dimensions. As $\rho$ becomes more negative, seesaw experimentation becomes more likely, reflecting situations where improvements in the primary dimension are increasingly linked to negative effects in the secondary dimension. While one might intuitively expect seesaw experimentation to occur only with negative correlation, our analysis reveals that this is not true. Because $(2\alpha M(\alpha) - 1)$ monotonically increases from $-1$ to $1$ for $\alpha \in [0,\infty)$, our sufficient condition indicates that seesaw experimentation can emerge even with positive correlation. This counterintuitive result occurs, for example, when $|\mu|$ is large, meaning the average negative effect on the secondary dimension is sufficiently strong to override the positive correlation between dimensions.

How to deal with seesaw experimentation? The traditional approach relies on inter-departmental coordination. For instance, customer satisfaction executives might veto an A/B test that boosts revenue but diminishes customer experience. However, this approach faces several challenges. The relative importance of different dimensions often involves subjective judgment, and outcomes can vary depending on individual executives' perspectives and personalities. Moreover, when companies carry out thousands of A/B tests, the sheer volume of experiments and the difficulty in measuring secondary dimension effects make coordination costs prohibitively high.

Below, we propose a coordination-free solution: the implementation of a positive hurdle rate $z$. This approach requires that any innovation must demonstrate a benefit exceeding $z$ in the primary dimension before adoption, effectively raising the bar for implementation without requiring complex coordination across departments.

When a positive hurdle $z$ is implemented, the firm's long-run average performance becomes:
\begin{align}
 \mathbb{E}[D(U + V)]
={}& \mathbb{P}(A=u) \mathbb{P}(U>z) \mathbb{E}[U+V | U>z]
+ \mathbb{P}(A=v) \mathbb{P}(V>z) \mathbb{E}[U+V | V>z] \nonumber \\
\coloneqq{}& f(z). \label{eq:f(z)}
\end{align}

\begin{proposition}\label{prop:opt-sym}
Suppose Assumptions~\ref{assump:iid} and \ref{assump:normal} hold and $(U, V)$ follows a  symmetric distribution.  If $\rho \in (-1, 1)$, then $f'(0) > 0 $, $f(z)$ attains its maximum at
$z^*$ and $f(z^*)>0$, where
\[z^* = \frac{\rho - 1}{\rho + 1}\mu.\]
\end{proposition}

Proposition \ref{prop:opt-sym} demonstrates that implementing a positive hurdle rate is strictly beneficial to the firm. The intuition is straightforward: when the firm sets a hurdle rate slightly above zero, there are two conflicting effects. First, there is a small loss in the primary dimension because some marginally beneficial innovations are not implemented. Note that the foregone benefits are small for these innovations. Second, there is a significant gain in the secondary dimension because the firm avoids implementing innovations that would have negative effects (in expectation) there. By setting a positive hurdle rate, the firm screens out potential innovations that might have small positive benefits in the primary dimension but large negative effects in the secondary dimension. This makes a positive hurdle rate beneficial.

This proposition also determines the optimal hurdle rate and reveals two key relationships. First, the optimal hurdle rate is proportionally related to $\mu$ (the mean value in the secondary dimension). When $\mu$ is lower, the firm sets a higher hurdle rate. This makes intuitive sense, as the firm wants to be more selective when secondary outcomes are likely to be less favorable.

Second, the proposition links the optimal hurdle rate with the correlation $\rho$ between primary and secondary outcomes. Note that when $\rho = 0$ (no correlation), the optimal hurdle rate equals $-\mu$, which is the expected negative impact in the secondary dimension. This arises because, with no correlation, $-\mu$ is the average negative secondary effect. When $\rho$ approaches $1$ (perfect positive correlation), the optimal hurdle rate approaches zero. This is because with perfect positive correlation, any innovation that is good for the primary dimension will also be good for the secondary dimension, eliminating the need for a positive hurdle. Finally, when $\rho$ approaches $-1$ (perfect negative correlation), the optimal hurdle rate approaches infinity. This extreme case occurs because perfect negative correlation means that any positive outcome in the primary dimension necessarily creates an equally large negative outcome in the secondary dimension, making the overall gain from innovation small.

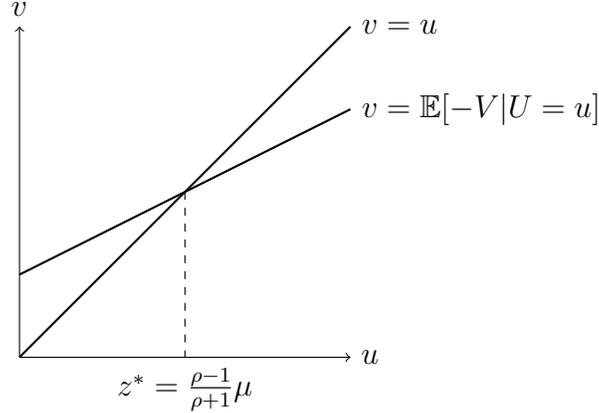
\begin{figure}[ht]
\begin{center}
\begin{tikzpicture}[scale=2.2]
    \draw[->] (0,0) -- (2,0) node[right] {$u$};
    \draw[->] (0,0) -- (0,2) node[above] {$v$};

    \draw[thick] (0,0) -- (2,2) node[right] {$v=u$};

    \draw[thick] (0,0.5) -- (2,1.5) node[right] {$v = \mathbb{E}[-V|U=u]$};

    \draw[dashed] (1,1) -- (1,0) node[below] {\( z^* = \frac{\rho-1}{\rho+1} \mu \)};
\end{tikzpicture}
\end{center}
\caption{Negative Externality and Optimal Hurdle Rate}\label{fig:z-star}
\end{figure}

The optimal hurdle rate is set by a basic economic principle: it should be where the positive gain from an innovation in the primary dimension equals the negative externality on the secondary dimension. At the optimal hurdle rate, the externality is internalized. Under the symmetric bivariate normal assumption, the negative externality is $\mathbb{E}[-V|U=u] = -\rho u  - (1-\rho)\mu$. As shown in Figure \ref{fig:z-star}, the gain in the primary dimension exceeds the expected negative externality only if it surpasses the optimal threshold $z^*=\frac{1-\rho}{1+\rho}\mu$.

Finally, note that the optimal hurdle rate in Proposition \ref{prop:opt-sym} maximizes the firm's long-term performance.
A firm that also prioritizes short-term objectives should set a lower hurdle rate, accepting more innovations despite their potential negative secondary effects.
This reflects the classic trade-off between short-term gains and long-term losses.

\begin{remark}[Asymmetry]
For asymmetric performance measurements, the firm can apply the same principle as in the symmetric case: set dimension-specific hurdle rates where primary dimension gains exactly offset secondary dimension externalities.
The optimal hurdle rate depends on both dimensions' parameters: notably, with a negative correlation, it decreases as the mean performance in the primary dimension increases, since more gains in one dimension reduce negative externalities in the other.
\end{remark}

\begin{remark}[Multidimensionality]
Our analysis extends to firms with multidimensional performance measurements.
As the number of dimensions increases, seesaw experimentation becomes more prevalent and optimal hurdle rates increase.
This follows from the increased likelihood  of negative spillovers across dimensions, thereby requiring stricter innovation adoption criteria.
\end{remark}

\begin{remark}[Fat Tails]
The study of \cite{azevedo2019abtesting} shows that A/B test outcomes often exhibit fat tails that normal distributions fail to capture.
We extend our analysis to bivariate $t$ distributions, which generalize normal distributions through a tail-thickness parameter, with lower values producing fatter tails.
Under symmetric bivariate $t$ distributions, seesaw experimentation becomes less common with fatter tails, which increase upside potential and thus make negative spillovers less likely to outweigh  gains in the primary dimension.
The optimal hurdle rate, however, remains unchanged from the normal model.
This is because both models yield identical expected externality expressions, and the optimal hurdle rate balances gains in the primary dimension against negative externalities in the secondary dimension (Figure~\ref{fig:z-star}).
\end{remark}

\section{Conclusion}

Our paper examines A/B testing in the presence of spillover effects. Building on a bivariate normal framework, we identify conditions that give rise to seesaw experimentation—where continuous adoption of successful innovations leads to performance decline. We show that this phenomenon is more likely to occur when the signal-to-noise ratio is large and correlations across dimensions are more negative. We also show how implementing appropriate hurdle rates can enhance firm performance, and derive optimal thresholds within our analytical framework. At the optimal threshold, the positive gain is equal to the negative externality it imposes.

While our results are theoretical in nature, they have practical implications: when innovations have multi-dimensional impacts, firms should move away from zero-threshold A/B testing. The interconnected nature of modern organizations demands a more systematic approach to experimentation—one that explicitly accounts for these cross-dimensional externalities. Setting optimal hurdle rates in practice requires careful measurement of innovation outcomes across dimensions, including their correlations. For this purpose, firms can leverage their historical A/B test data, which provides a rich source of information about both direct effects and spillovers.

\bibliographystyle{apalike}
\bibliography{ref}

\newpage

\begin{appendix}
\numberwithin{equation}{section}

\bigskip
\begin{center}
    \LARGE {\textbf{Appendices}}
\end{center}

\section{Extensions}\label{sec:ext}

This section extends the results of Section \ref{section: symmetric} in three directions: asymmetry, multidimensionality, and fat tails. For clarity of managerial insights, we examine each extension separately rather than analyzing a fully general model combining all features.

\subsection{Asymmetry}

Consider a general bivariate normal distribution without the symmetry assumptions $\mu_u = \mu_v$ and $\sigma_u = \sigma_v$. Due to this asymmetry between dimensions, the firm sets distinct hurdle rates $z_u$ and $z_v$. The firm's long-run average performance then becomes:
\begin{align}\label{eq:perf-asymp}
 & \mathbb{E}[D(U + V)] \nonumber\\
& = {} \mathbb{P}(A=u) \mathbb{P}(U>z_u) \mathbb{E}[U+V | U>z_u]
+ \mathbb{P}(A=v) \mathbb{P}(V>z_v) \mathbb{E}[U+V | V>z_v] \nonumber \\
& \coloneqq g(z_u, z_v).
\end{align}

\begin{proposition}\label{prop:seesaw-asym}
Suppose Assumptions~\ref{assump:iid} and \ref{assump:normal} hold.
If $-1 \leq \rho < \min(\rho_1, \rho_2)$, then $g(0,0) <0$, where
\begin{align*}
\rho_1 = {}&
 \left[ \left(\frac{-\mu_u}{\sigma_u}\right)M\left(\frac{ - \mu_u}{\sigma_u} \right) \left(1 + \frac{\mu_v}{\mu_u} \right)  - 1 \right]\frac{\sigma_u}{\sigma_v}, \\
\rho_2 ={}&
 \left[ \left(\frac{-\mu_v}{\sigma_v}\right)M\left(\frac{ - \mu_v}{\sigma_v} \right) \left(1 + \frac{\mu_u}{\mu_v} \right)  - 1  \right]\frac{\sigma_v}{\sigma_u}.
\end{align*}

\end{proposition}

\begin{proposition}\label{prop:opt-asym}
Suppose Assumptions~\ref{assump:iid} and \ref{assump:normal} hold.
If
\[
- \min\left(\frac{\sigma_v}{\sigma_u}, \frac{\sigma_u}{\sigma_v}\right)
< \rho < \min\left(\frac{\mu_v/\sigma_v}{\mu_u/
\sigma_u}, \frac{\mu_u/
\sigma_u}{\mu_v/\sigma_v}\right),
\]
then $\nabla g(0, 0) > 0$,
$g(z_u, z_v)$ attains its maximum
at $(z_u^*, z_v^*)$, and $g(z_u^*, z_v^*) > 0$ where
\[z_u^* = \frac{\rho \mu_u / \sigma_u - \mu_v / \sigma_v}{\rho/ \sigma_u + 1 / \sigma_v}
\quad\mbox{and}\quad
z_v^* = \frac{\rho \mu_v / \sigma_v - \mu_u / \sigma_u}{\rho/ \sigma_v + 1 / \sigma_u}.\]
\end{proposition}

The conditions in Proposition \ref{prop:seesaw-asym} reduce to that in Proposition \ref{prop:seesaw-sym} when the means and variances of the two dimensions are equal. Proposition \ref{prop:opt-asym} derives the optimal hurdle rates for both dimensions. The optimal hurdle rate for each dimension follows the same principle as in the symmetric case: it is set to the value at which the positive gain from adopting an innovation exactly offsets its negative externality on the other dimension.

The optimal hurdle rate increases when either the mean of the secondary dimension is lower or when the correlation coefficient is lower, following the same logic as in the symmetric case since both conditions lead to greater negative externalities.

In addition, the optimal hurdle rate also depends on the mean of the primary dimension, except when the correlation coefficient $\rho$ equals zero. In the special case where $\rho = 0$, the optimal hurdle rate in one dimension exactly offsets the negative externality imposed on the other (e.g., $z_u^* = -\mu_v$). When $\rho$ is negative, the optimal hurdle rate
$z_u^*$ decreases as $\mu_u$ increases, because a higher mean in the primary dimension reduces the negative externality, leading to a higher optimal hurdle rate.\footnote{Under the bivariate normal assumption, the negative externality is $\mathbb{E}[-V|U]= \rho(\sigma_v/\sigma_u)(\mu_u - U) - \mu_v$. As a result, when $\rho<0$, an increase in $\mu_u$ reduces the negative externality.} Similarly, when $\rho$ is positive, $z_u^*$ increases in $\mu_u$.

\subsection{Multidimensionality}

Now suppose the firm measures performance across $n$ dimensions.
Each period, it identifies a dimension $A$ as its strategic priority, evaluates a potential innovation designed for dimension $A$ through an A/B test, and adopts the innovation if its effect on dimension $A$ exceeds a threshold $z$.
Let $X_i$ denote the innovation's effect on dimension $i$ ($i=1,\ldots,n$), although only the effect on the primary dimension $A$ is measured.
Define the adoption decision as $D = \mathbb{I}(X_A > z)$.
The long-run average of the firm's overall performance is then:
\begin{equation}\label{eq:perf-multi}
\mathbb{E}\left[D\sum_{i=1}^n X_i\right] = \sum_{i=1}^n \mathbb{P}(A=i)\mathbb{P}(X_j > z)\mathbb{E}\left[D\sum_{i=1}^n X_i\bigg| X_j>z\right] \coloneqq h(z).
\end{equation}

\begin{assumption}\label{assump:MVN}
\begin{enumerate}[label=(\roman*), noitemsep, topsep=0pt]
   \item $(X_1,\ldots, X_n)$ follows a multivariate normal distribution with mean vector $m$ and covariance matrix $\Sigma$:
\[m = \begin{pmatrix} \mu \\ \mu \\ \vdots \\ \mu \end{pmatrix}\quad \mbox{and}\quad \Sigma =  \begin{pmatrix}\sigma^2 & \rho \sigma^2 & \cdots & \rho \sigma^2  \\ \rho \sigma^2 &  \sigma^2 & \cdots & \rho \sigma^2 \\ \vdots & \vdots & \ddots & \vdots \\ \rho \sigma^2  & \rho \sigma^2 & \cdots & \sigma^2\end{pmatrix}. \]
\item $\mu < 0$.
\end{enumerate}
\end{assumption}

\begin{proposition}\label{prop:seesaw-mult}
Suppose Assumptions \ref{assump:iid} and \ref{assump:MVN} hold.
Then, $h(0) < 0$ if
\[
-\frac{1}{n-1} \leq \rho < \left(\frac{n}{n-1}\right) \left(\frac{-\mu}{\sigma} \right) M\left(\frac{-\mu}{\sigma} \right) - \frac{1}{n-1}.
\]
\end{proposition}

\begin{proposition}\label{prop:opt-mult}
Suppose Assumptions \ref{assump:iid} and \ref{assump:MVN} hold.
If $-1/(n-1) < \rho < 1 $, then $h'(0) > 0$,
$h(z)$ attains its maximum at $z^*$, and $h(z^*)>0$, where
\[z^* = \frac{(n-1)(\rho - 1)}{ (n-1)\rho + 1} \mu .\]
\end{proposition}

When $n=2$, the expressions in Propositions \ref{prop:seesaw-mult} and \ref{prop:opt-mult} reduce to those in our bivariate-normal analysis. In Proposition \ref{prop:seesaw-mult}, the lower bound $-1/(n-1)$ is the lowest feasible correlation coefficient in a multivariate normal distribution. This indicates that seesaw experimentation emerges when A/B test outcomes across different dimensions are sufficiently negatively correlated, consistent with our earlier findings. Proposition \ref{prop:opt-mult} shows that the optimal hurdle rate increases in $n$. This reflects an intuitive relationship: more dimensions create more negative externalities, leading to a higher optimal threshold for innovation adoption.

\subsection{Fat Tails}

\cite{azevedo2019abtesting} found that A/B test results frequently display fat tails that cannot be adequately modeled by normal distributions.
Building on this insight, we extend our analysis to bivariate $t$ distributions.
This class of distributions extends bivariate normal distributions by adding a parameter that controls tail thickness. Unlike normal distributions, their tail probabilities decay polynomially, with the added parameter determining the decay rate.\footnote{While Pareto distributions also model heavy tails effectively, their bivariate extensions have limited flexibility: they are not defined over the entire real plane, and their correlation structure is constrained by the tail thickness parameter rather than being independently specified  \citep{Arnold15}.}

\begin{assumption}\label{assump:t}
\begin{enumerate}[label=(\roman*), noitemsep, topsep=0pt]
   \item $(U, V)$ follows a bivariate $t$ distribution with mean vector $m$, scale matrix $\Sigma$, and degrees of freedom $\delta$:
\[m = \begin{pmatrix} \mu \\ \mu \end{pmatrix}\quad \mbox{and}\quad \Sigma =  \begin{pmatrix}\sigma^2 & \rho \sigma^2 \\  \rho \sigma^2  & \sigma^2\end{pmatrix}. \]
\item $\mu < 0$ and $\delta>2$.
\end{enumerate}
\end{assumption}

The parameter $\delta$ controls the tail thickness of the distribution, with smaller values of $\delta$ corresponding to heavier tails.
When $\delta \leq 2$, the distribution lacks finite second moments, making covariance and correlation undefined.
For $\delta > 2$, the scale matrix $\Sigma$ equals the covariance matrix of $(U, V)$, where
$\sigma^2$ represents the marginal variance and $\rho$ the correlation.
As $\delta \to\infty$, the distribution converges to a bivariate normal with the same mean vector and covariance matrix, eliminating the fat tails.

\begin{proposition}\label{prop:seesaw-sym-t}
Suppose Assumptions~\ref{assump:iid} and \ref{assump:t} hold.
If
\begin{equation}\label{eq:upper-bound-t}
    -1 \leq \rho < 2\left(\frac{-\mu}{\sigma}\right) W\left(\frac{-\mu}{\sigma}\right) - 1,
\end{equation}
then $f(0) <0$, where
\[
W(\alpha) = \left(\frac{\delta-2}{\delta}\right) \left(\frac{1 - T_{\delta}(\alpha; 0, 1)}{t_{\delta-2}\left(\alpha; 0, \frac{\delta}{\delta-2}\right)}\right),
\]
$t_\delta(\alpha; \mu, \sigma^2)$ and $T_\delta(\alpha; \mu, \sigma^2)$ denote the probability density function and cumulative distribution function, respectively, of a univariate $t$ distribution with mean $\mu$, variance $\sigma^2$, and degrees of freedom $\delta$.
\end{proposition}

\begin{figure}[ht]
    \centering
    \includegraphics[width=0.70\textwidth]{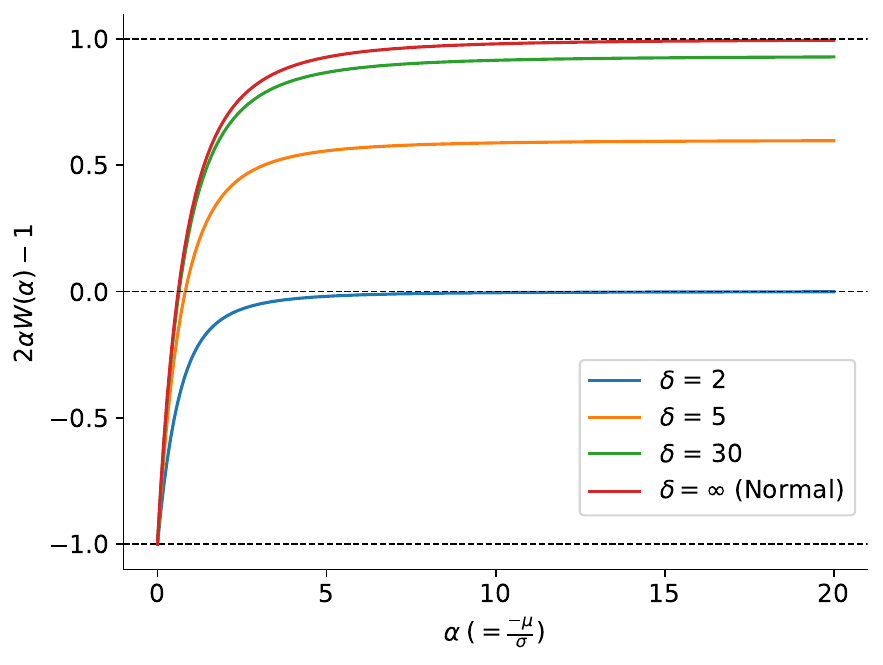}
    \caption{Maximum Correlation Leading to Seesaw Experimentation}
    \label{fig:upper-bound}
\end{figure}

In equation \eqref{eq:upper-bound-t}, the maximum correlation that permits seesaw experimentation is $2\alpha W(\alpha) - 1$, where $\alpha = -\mu/\sigma$.
Figure \ref{fig:upper-bound} illustrates this threshold as a function of $\alpha$ for different values of $\delta$.
As $\delta$ increases, this threshold increases for any given $\alpha$.
This indicates that, holding marginal mean ($\mu$) and variance ($\sigma^2$) constant,
lighter tails expand the range of correlations that permit seesaw experimentation, making such phenomenon more likely to occur. This result has an intuitive explanation: with lighter tails, experimentation offers more limited upside potential. Consequently, the negative spillover effects in the secondary dimension are more likely to outweigh the positive gains in the primary dimension.

\begin{remark}
The standard $t$ distribution with $\delta$ degrees of freedom converges to the standard normal distribution as $\delta\to\infty$.
Thus,
\[W(\alpha) = \left(\frac{\delta-2}{\delta}\right) \left(\frac{1 - T_{\delta}(\alpha; 0, 1)}{t_{\delta-2}\left(\alpha; 0, \frac{\delta}{\delta-2}\right)}\right)  \to
\frac{1 - T_{\infty}(\alpha; 0, 1)}{t_{\infty}(\alpha; 0, 1)}  =  \frac{1 - \Phi(\alpha)}{\phi(\alpha)} = M(\alpha).\]
Hence,
the upper bound in equation \eqref{eq:upper-bound-t} for the bivariate $t$ distribution converges to the upper bound in
\eqref{eq:upper-bound-normal} for the base model (bivariate normal distribution).
\end{remark}

\begin{proposition}\label{prop:opt-sym-t}
Suppose Assumptions~\ref{assump:iid} and \ref{assump:t} hold.
If $\rho \in (-1, 1)$, then $f'(0) > 0 $, $f(z)$ attains its maximum at
$z^*$ and $f(z^*)>0$, where
\[z^* = \frac{\rho-1}{\rho+1}\mu.\]
\end{proposition}
Note that the choice of the optimal hurdle rate is the same as that in the normal distribution.
The underlying intuition, illustrated in Figure \ref{fig:z-star}, holds true: the optimal hurdle rate equates the expected negative externality in the secondary dimension with the gain in the primary dimension.
Both bivariate normal and bivariate $t$ distributions share the same expression for the expected negative externality: $\mathbb{E}[-V|U=u] = -\rho \mu - (1-\rho) \mu$.

\section{Technical Proofs}

\subsection{Proofs for Bivariate Normal Distribution}

\begin{lemma}\label{lemma:Mills}
    $k(\alpha) \coloneqq \alpha M(\alpha)$ is strictly increasing in $\alpha \in \mathbb{R}$, $k(0) = 0$, and $\lim\limits_{\alpha\to\infty}k(\alpha)= 1$.
\end{lemma}
\begin{proof}[Proof of Lemma~\ref{lemma:Mills}]
Note: $k(\alpha) = \alpha(1 - \Phi(\alpha)) / \phi(\alpha) $ and $\phi'(\alpha) = (- \alpha) \phi(\alpha) $.
Hence,
\begin{align*}
    k'(\alpha) = \frac{(1 - \Phi(\alpha))(1+\alpha^2) - \alpha \phi(\alpha)}{\phi(\alpha)}.
\end{align*}
Because $\phi(\alpha)>0$ and $\Phi(\alpha) \in (0,1)$ for all $\alpha$,
we have $k'(\alpha)>0$ for all $\alpha \leq 0$.
Moreover,
it is known that the Mills ratio satisfies
\[\frac{1 - \Phi(\alpha)}{\alpha} > \frac{\alpha}{1+\alpha^2},\quad \mbox{for all }\alpha>0;\]
see, e.g., \cite{GasullUtzet14}.
Hence, $k'(\alpha) >0 $ for all $\alpha>0$.

Finally, the limit of $k(\alpha)$ can be calculated via L'H\^{o}pital's rule:
\end{proof}

\begin{lemma}\label{lemma:long-run-avg-z}
    Let $g(z_u, z_v)$ be as defined in \eqref{eq:perf-asymp}.
    Then,
\begin{equation}\label{eq:long-run-avg-z}
\begin{aligned}
g(z_u, z_v) = {} & p_u \left[(\mu_u + \mu_v ) \left(1-\Phi\left(\frac{z_u - \mu_u}{\sigma_u} \right) \right) + \left(\sigma_u + \rho \sigma_v  \right) \phi\left(\frac{z_u - \mu_u}{\sigma_u}\right) \right] \\
& + p_v \left[(\mu_u + \mu_v ) \left(1-\Phi\left( \frac{z_v - \mu_v}{\sigma_v} \right) \right) + \left(\rho \sigma_u +  \sigma_v  \right) \phi\left(\frac{z_v - \mu_v}{\sigma_v}\right) \right].
\end{aligned}
\end{equation}
\end{lemma}

\begin{proof}[Proof of Lemma~\ref{lemma:long-run-avg-z}]
We first note that
\begin{align*}
\mathbb{E}[D(U + V)]
={}& p_u \mathbb{E}[U+V | U>z_u] \mathbb{P}(U>z_u) + p_v \mathbb{E}[U+V | V>z_v]  \mathbb{P}(V>z_v).
\end{align*}
The expression \eqref{eq:long-run-avg-z}  can be derived by using the following properties of the truncated normal distribution \cite[p.~112--113]{johnson72distribution}:
for any $z$,
\begin{equation*}\label{eq:mills}
\mathbb{E}[U|U>z] = \mu_u +  \sigma_u \lambda\left(\frac{z - \mu_u}{\sigma_u}\right) \quad\mbox{and}\quad
    \mathbb{E}[V|U>z] = \mu_v + \rho \sigma_v \lambda\left(\frac{z - \mu_u}{\sigma_u}\right),
\end{equation*}
where $\lambda(\alpha) \coloneqq 1/M(\alpha)$ is the inverse Mills ratio of the standard normal distribution.
\end{proof}

\begin{proof}[Proof of Proposition~\ref{prop:seesaw-sym}]
The proof follows from $f(0) = g(0, 0)$ and Lemma~\ref{lemma:long-run-avg-z} through direct calculations.
\end{proof}

\begin{proof}[Proof of Proposition~\ref{prop:opt-sym}]

Note that
\begin{align*}
f(z) ={}& g(z, z)
=  2\mu \left(1-\Phi\left( \frac{z-\mu}{\sigma} \right) \right) + \sigma (1+\rho)    \phi\left(\frac{z-\mu}{\sigma}\right), \\
    f'(z) ={}& -\frac{1}{\sigma} \phi\left(\frac{z-\mu}{\sigma}\right)\left[2\mu + (1+\rho) (z-\mu) \right], \\
    f''(z) ={}& -\frac{1}{\sigma} \phi\left(\frac{z-\mu}{\sigma}\right)\left[ \left(-\frac{1}{\sigma}\right) \left( \frac{z-\mu}{\sigma}\right) \left(2\mu + (1+\rho) (z-\mu) \right) + 1 + \rho\right].
\end{align*}

Hence, if $-1 < \rho < 1 $, then $f'(0) > 0$  and $f'(z) = 0$ has a unique solution
\[z^* = \frac{\rho - 1}{ \rho + 1} \mu;\]
moreover,
\[f''(z^*) = -\frac{1}{\sigma} \phi\left(\frac{-2\mu}{\sigma(1+\rho)}\right) (1 + \rho) < 0,\]
which implies $z^*$ the a maximizer of $f(z)$.
Moreover,
   \begin{align*}
       f(z^*) ={}& \sigma (1+\rho) \phi\left(\frac{z^* - \mu}{\sigma}\right)
       \left[ \frac{2\mu}{\sigma(1+\rho)} M\left(\frac{z^* - \mu}{\sigma}\right) + 1 \right] \\
       ={}& \sigma (1+\rho) \phi\left(\frac{z^* - \mu}{\sigma}\right)
       \left[ - \left(\frac{z^* - \mu}{\sigma}\right) M\left(\frac{z^* - \mu}{\sigma}\right) + 1 \right] > 0,
   \end{align*}
where the positivity follows from $\alpha M(\alpha) < 1$ by Lemma~\ref{lemma:Mills}.
\end{proof}

\begin{proof}[Proof of Proposition~\ref{prop:seesaw-asym}]
By equation~\eqref{eq:long-run-avg-z}, a sufficient condition for $g(0,0) <0 $ is
\begin{equation}
    \left\{
    \begin{array}{l}
    \displaystyle (\mu_u + \mu_v ) \left(1-\Phi\left(\frac{ - \mu_u}{\sigma_u} \right) \right) + \left(\sigma_u + \rho \sigma_v  \right) \phi\left(\frac{ - \mu_u}{\sigma_u}\right) < 0, \\[1em]
     \displaystyle  (\mu_u + \mu_v ) \left(1-\Phi\left( \frac{ - \mu_v}{\sigma_v} \right) \right) + \left(\rho \sigma_u +  \sigma_v  \right) \phi\left(\frac{ - \mu_v}{\sigma_v}\right) < 0.
    \end{array}
    \right.
\end{equation}
By straightforward calculations,
this condition is the same as $\rho < \min(\rho_1,\rho_2)$, where
\begin{align*}
\rho_1 = {}&
 \left[ \left(\frac{-\mu_u}{\sigma_u}\right)M\left(\frac{ - \mu_u}{\sigma_u} \right) \left(1 + \frac{\mu_v}{\mu_u} \right)  - 1 \right]\frac{\sigma_u}{\sigma_v}, \\
\rho_2 ={}&
 \left[ \left(\frac{-\mu_v}{\sigma_v}\right)M\left(\frac{ - \mu_v}{\sigma_v} \right) \left(1 + \frac{\mu_u}{\mu_v} \right)  - 1  \right]\frac{\sigma_v}{\sigma_u}.
\end{align*}
\end{proof}

\begin{proof}[Proof of Proposition~\ref{prop:opt-asym}]

Note that
\begin{align*}
\frac{\partial g(z_u, z_v)}{\partial z_u}  = {}& p_u \left(-\frac{1}{\sigma_u}\right) \phi  \left(\frac{z-\mu_u}{\sigma_u}\right) \left[(\mu_u + \mu_v) + (\sigma_u + \rho \sigma_v) \left(\frac{z-\mu_u}{\sigma_u}\right)\right], \\[1em]
\frac{\partial g(z_u, z_v)}{\partial z_v}  = {}& p_v \left(-\frac{1}{\sigma_v}\right) \phi  \left(\frac{z-\mu_v}{\sigma_v}\right) \left[(\mu_u + \mu_v) + (\rho\sigma_u + \sigma_v) \left(\frac{z-\mu_v}{\sigma_v}\right)\right].
\end{align*}
Hence, $\nabla g(0, 0) > 0$ if and only if
\[
(\mu_u + \mu_v) + (\sigma_u + \rho \sigma_v) \left(\frac{-\mu_u}{\sigma_u}\right) < 0 \quad\mbox{and}\quad
 (\mu_u + \mu_v) + (\rho\sigma_u + \sigma_v) \left(\frac{-\mu_v}{\sigma_v}\right) < 0,
\]
which is equivalent to
\[\rho < \frac{\mu_v/\sigma_v}{\mu_u/
\sigma_u}\quad\mbox{and}\quad \rho < \frac{\mu_u/
\sigma_u}{\mu_v/\sigma_v}.\]

If $\rho > - \min(\sigma_v / \sigma_u, \sigma_u / \sigma_v)$, then $\nabla g(z_u, z_v) = 0$ has a unique solution
\[z_u^* = \frac{\rho \mu_u / \sigma_u - \mu_v / \sigma_v}{\rho/ \sigma_u + 1 / \sigma_v}
\quad\mbox{and}\quad
z_v^* = \frac{\rho \mu_v / \sigma_v - \mu_u / \sigma_u}{\rho/ \sigma_v + 1 / \sigma_u};\]
moreover,
\begin{align*}
\frac{\partial^2 g(z_u^*, z_v^*)}{\partial z_u^2}  = {}& p_u\left(-\frac{1}{\sigma_u}\right) \phi \left(\frac{-(\mu_u + \mu_v)}{ \sigma_u +  \rho \sigma_v} \right)\left(1 + \frac{\sigma_v}{\sigma_u} \rho\right) < 0, \\[1em]
\frac{\partial^2 g(z_u^*, z_v^*)}{\partial z_v^2}  = {}& p_v\left(-\frac{1}{\sigma_v}\right) \phi \left(\frac{-(\mu_u + \mu_v)}{\rho \sigma_u +  \sigma_v} \right)\left(1 + \frac{\sigma_u}{\sigma_v} \rho\right) <0,  \\[1em]
\frac{\partial^2 g(z_u^*, z_v^*)}{\partial z_u \partial z_v}  = {}&  \frac{\partial^2 g(z_u^*, z_v^*)}{\partial z_v \partial z_u} = 0.
\end{align*}
Thus, $\nabla^2 g(z_u^*, z_v^*) $ is negative definite, implying $(z_u^*, z_v^*)$ is the maximizer of $g(z_u, z_v)$.
Moreover,
\begin{align*}
    g(z_u^*, z_v^*) ={}& p_u (\sigma_u + \rho\sigma_v) \phi\left(\frac{z_u^* - \mu_u}{\sigma_u}\right) \left[\left(\frac{\mu_u + \mu_v}{\sigma_u + \rho\sigma_v} \right)M\left(\frac{z_u^* - \mu_u}{\sigma_u}\right) + 1  \right] \\
    & + p_v (\sigma_u + \rho\sigma_v) \phi\left(\frac{z_v^* - \mu_v}{\sigma_v}\right) \left[\left(\frac{\mu_u + \mu_v}{\rho \sigma_u + \sigma_v} \right)M\left(\frac{z_v^* - \mu_v}{\sigma_v}\right) + 1  \right] \\
    ={}& p_u (\sigma_u + \rho\sigma_v) \phi\left(\frac{z_u^* - \mu_u}{\sigma_u}\right) \left[- \left(\frac{z_u^* - \mu_u}{\sigma_u}\right) M\left(\frac{z_u^* - \mu_u}{\sigma_u}\right) + 1  \right] \\
    & + p_v (\sigma_u + \rho\sigma_v) \phi\left(\frac{z_v^* - \mu_v}{\sigma_v}\right) \left[- \left(\frac{z_v^* - \mu_v}{\sigma_v}\right)M\left(\frac{z_v^* - \mu_v}{\sigma_v}\right) + 1  \right] > 0,
\end{align*}
where the positivity follows from Lemma~\ref{lemma:Mills}, which implies $\alpha M(\alpha) < 1$, and the assumption that $\rho > - \min(\sigma_v / \sigma_u, \sigma_u / \sigma_v)$, which implies $\sigma_u + \rho \sigma_v>0$ and $\rho\sigma_mu + \sigma_v > 0$.
\end{proof}

\subsection{Proofs for Multivariate Normal Distribution}

\begin{lemma}\label{lemma:long-run-avg-z-multi}
Let $h(z)$ be as defined in \eqref{eq:perf-multi}. Then,
\[h(z) = \mathbb{E}\left[D\sum_{i=1}^n X_i\right]
=  n \mu \left(1-\Phi\left( \frac{z-\mu}{\sigma} \right) \right) + \sigma [1+(n-1)\rho]    \phi\left(\frac{z-\mu}{\sigma}\right).\]
\end{lemma}
\begin{proof}[Proof of Lemma~\ref{lemma:long-run-avg-z-multi}]
The proof follows similar calculations to those in Lemma~\ref{lemma:long-run-avg-z}.
\end{proof}

\begin{proof}[Proof of Proposition~\ref{prop:seesaw-mult}]
The proof follows from Lemma~\ref{lemma:long-run-avg-z-multi} through direct calculations.
\end{proof}

\begin{proof}[Proof of Proposition~\ref{prop:opt-mult}]
Note that
\begin{align*}
    h'(z) ={}& -\frac{1}{\sigma} \phi\left(\frac{z-\mu}{\sigma}\right)\left[n\mu + [1+(n-1)\rho] (z-\mu) \right] \\
    h''(z) ={}& -\frac{1}{\sigma} \phi\left(\frac{z-\mu}{\sigma}\right)\left[ \left(-\frac{1}{\sigma}\right) \left( \frac{z-\mu}{\sigma}\right) \left(n\mu + [1+(n-1)\rho] (z-\mu) \right) + 1 + (n-1)\rho\right] .
\end{align*}

Hence, if $-1/(n-1) < \rho  < 1$, then $h'(z) = 0$ has a unique solution
\[z^* = \frac{(n-1)(\rho - 1)}{ (n-1)\rho + 1 } \mu ;\]
moreover,
\[h''(z^*) = -\frac{1}{\sigma} \phi\left(\frac{-n\mu}{\sigma[1+(n-1)\rho]}\right) [1 + (n-1)\rho] < 0,\]
which implies $z^*$ is the maximizer of $h(z)$.
In addition,
\begin{align*}
    h(z^*) ={}& \sigma [1+(n-1)\rho)] \phi\left(\frac{z^* - \mu}{\sigma}\right) \left[\frac{n\mu}{\sigma[1+(n-1)\rho]} M\left(\frac{z^* - \mu}{\sigma}\right) +  1 \right] \\
    ={}& \sigma [1+(n-1)\rho)] \phi\left(\frac{z^* - \mu}{\sigma}\right) \left[-\left(\frac{z^* - \mu}{\sigma}\right) M\left(\frac{z^* - \mu}{\sigma}\right) +  1 \right] > 0,
\end{align*}
where the positivity follows from Lemma~\ref{lemma:Mills} and the assumption that $\rho > -1/(n-1)$.
\end{proof}

\subsection{Proofs for Bivariate $t$ Distribution}

\begin{lemma}\label{lemma:f(z)-t}
Suppose $(U, V)$ follows a symmetric bivariate $t$ distribution. Then,
\begin{align*}
f(z) ={}&  2\mu \left(1-T_\delta\left( \frac{z-\mu}{\sigma};0, 1  \right)\right) + \sigma (1+\rho)   \left(\frac{\delta}{\delta-2} \right) t_{\delta-2}\left(\frac{z-\mu}{\sigma}; 0, \frac{\delta}{\delta-2}\right).
\end{align*}
\end{lemma}

\begin{proof}[Proof of Lemma \ref{lemma:f(z)-t}]
Each marginal ($U$ or $V$) follows a univariate $t$ distribution with mean $\mu$, variance $\sigma^2$, and degrees of freedom $\delta>2$.
Then, $\mathbb{E}[U|U>z]$ is the expectation of a truncated $t$ distribution with truncation interval $(z, \infty)$.
Hence,
\begin{equation}\label{eq:truncated-t-expect}
    \mathbb{E}[U|U>z] = \mu + \sigma
\frac{\delta \Gamma\left(\frac{\delta+1}{2}\right)}{(\delta-1)\sqrt{\delta \pi}\Gamma\left(\frac{\delta}{2}\right) }\left(1 + \frac{(z-\mu)^2}{\sigma^2\delta}\right)^{-\frac{\delta-1}{2}}\frac{1}{1 - T_\delta\left(\frac{z - \mu}{\sigma}; 0, 1\right)},
\end{equation}
where $\Gamma$ denotes the gamma function;
see, e.g., \cite{HoLinChenWang12}.

Note that the $t$ density function $t_\delta(z; \mu, \sigma^2)$ is
\[
t_\delta(z; \mu, \sigma^2) = \frac{\Gamma\left(\frac{\delta+1}{2}\right)}{\sqrt{\sigma^2\delta \pi}\Gamma\left(\frac{\delta}{2}\right)}
\left(1 + \frac{(z-\mu)^2}{\sigma^2\delta}\right)^{-\frac{\delta+1}{2}}.
\]
Therefore,
\begin{align*}
    t_{\delta-2}\left(\frac{z-\mu}{\sigma};0,\frac{\delta}{\delta-2}\right) = {}&\frac{\Gamma(\frac{\delta-1}{2})}{\sqrt{\frac{\delta}{\delta-2} (\delta-2)\pi} \Gamma(\frac{\delta-2}{2})} \left(1+\frac{(\frac{z-\mu}{\sigma})^2}{ \frac{\delta}{\delta-2} (\delta-2)}\right)^{-\frac{\delta-1}{2}}  \\
    ={}  & \frac{\Gamma(\frac{\delta-1}{2})}{\sqrt{\delta \pi} \Gamma(\frac{\delta-2}{2})} \left(1+\frac{(z-\mu)^2}{\sigma^2 \delta}\right)^{-\frac{\delta-1}{2}} \\
    ={} & \frac{(\delta-2)\Gamma(\frac{\delta+1}{2}) }{(\delta-1)\sqrt{\delta \pi} \Gamma(\frac{\delta}{2}) } \left(1+\frac{(z-\mu)^2}{\sigma^2 \delta}\right)^{-\frac{\delta-1}{2}},
\end{align*}
where the last step follows from the gamma function's property $\Gamma(z+1) = z \Gamma(z)$.
Hence,
we can rewrite
\eqref{eq:truncated-t-expect} as
\begin{equation}\label{eq:t:E[U|U>z]}
\mathbb{E}[U|U>z] = \mu + \sigma \left(\frac{\delta}{\delta-2} \right)\left(\frac{t_{\delta-2}\left(\frac{z - \mu}{\sigma}; 0, \frac{\delta}{\delta-2}\right)}{1 - T_\delta\left(\frac{z - \mu}{\sigma}; 0, 1\right)}\right).
\end{equation}

The conditional distribution of $V$ given $U$ is a univariate $t$ distribution (see, e.g., \cite{Ding16}).
In particular,
$\mathbb{E}[V|U] = \mu + \rho (U - \mu)$.
Thus,
\begin{align}
\mathbb{E}[V |U > z] ={}& \mathbb{E}\bigl[\mathbb{E}[V|U]\big|U>z\bigr]
= \mu + \rho \bigl[\mathbb{E}[U|U>z] -\mu \bigr] \nonumber \\
={}& \mu + \rho \sigma \left(\frac{\delta}{\delta-2} \right)\left(\frac{t_{\delta-2}\left(\frac{z - \mu}{\sigma}; 0, \frac{\delta}{\delta-2}\right)}{1 - T_\delta\left(\frac{z - \mu}{\sigma}; 0, 1\right)}\right). \label{eq:t:E[V|U>z]}
\end{align}
Therefore,
\begin{align*}
f(z) ={}&  \mathbb{E}[D(U + V)]  \\
={}& \mathbb{P}(A=u) \mathbb{P}(U>z) \mathbb{E}[U+V | U>z]
+ \mathbb{P}(A=v) \mathbb{P}(V>z) \mathbb{E}[U+V | V>z] \\
={}& \mathbb{P}(U>z) \mathbb{E}[U+V | U>z] \\
={}& 2\mu \left(1 - T_\delta\left(\frac{z - \mu}{\sigma}; 0, 1\right) \right) + \sigma(1+\rho) \left(\frac{\delta}{\delta-2} \right) t_{\delta-2}\left(\frac{z-\mu}{\sigma}; 0, \frac{\delta}{\delta-2}\right),
\end{align*}
where the third equality holds because $\mathbb{E}[U + V |U > z] = \mathbb{E}[U + V |V > z]$ (symmetry), and the last equality follows from \eqref{eq:t:E[U|U>z]} and \eqref{eq:t:E[V|U>z]}.
\end{proof}

\begin{proof}[Proof of Proposition~\ref{prop:seesaw-sym-t}]
The proof follows from Lemma~\ref{lemma:long-run-avg-z} through direct calculations.
\end{proof}

\begin{proof}[Proof of Proposition \ref{prop:opt-sym-t}]
Direct calculations based on the expression of $f(z)$ in Lemma~\ref{lemma:long-run-avg-z} yield:
\begin{align*}
    f'(z) ={}& -\frac{1}{\sigma} t_\delta\left(\frac{z-\mu}{\sigma};0, 1\right) \left[2\mu + (1+\rho) (z-\mu) \right],
\end{align*}
where the following intermediate result is used
\begin{align*}
\frac{\dd}{\dd \alpha} t_{\delta-2}\left(\alpha; 0, \frac{\delta}{\delta-2}\right) = (-\alpha) \frac{\delta-2}{\delta} t_\delta(\alpha; 0, 1).
\end{align*}

Hence, if $-1 < \rho < 1 $, then $f'(0) > 0$ and  $f'(z)=0$  has a unique solution
\[z^* = \frac{\rho-1}{\rho+1}\mu.\]

Furthermore, we have
\begin{align*}
    f''(z) ={}  -\frac{1}{\sigma} \bigg[ & \frac{\dd}{\dd z} t_{\delta}\left(\frac{z-\mu}{\sigma}; 0, 1 \right) \left(2\mu + (1+\rho) (z-\mu) \right)  + t_{\delta}\left(\frac{z-\mu}{\sigma}; 0, 1 \right)  (1+\rho) \bigg].
\end{align*}
Thus,
\begin{align*}
    f''(z^*) = -\frac{1}{\sigma} t_{\delta}\left(\frac{z^*-\mu}{\sigma}; 0, 1 \right)  (1+\rho) < 0,
\end{align*}
proving $z^*$ is the maximum of $f(z)$.

Moreover,
\begin{align}
    f(z^*) = {}& 2\mu \left[1 - T_\delta\left(\frac{z^* - \mu}{\sigma}; 0, 1\right) \right] \left[ 1 + \frac{\sigma(1+\rho) }{2\mu} \left(\frac{\delta}{\delta-2} \right)\frac{t_{\delta-2}\left(\frac{z^*-\mu}{\sigma}; 0, \frac{\delta}{\delta-2}\right)}{1 - T_\delta\left(\frac{z^* - \mu}{\sigma}; 0, 1\right) }  \right]\nonumber \\
    ={}& 2\mu \left[1 - T_\delta\left(\frac{z^* - \mu}{\sigma}; 0, 1\right) \right] \left[ 1 + \frac{\sigma }{-(z^*-\mu)} \left(\frac{\delta}{\delta-2} \right)\frac{t_{\delta-2}\left(\frac{z^*-\mu}{\sigma}; 0, \frac{\delta}{\delta-2}\right)}{1 - T_\delta\left(\frac{z^* - \mu}{\sigma}; 0, 1\right) }  \right]. \label{eq:f(z^*)-t}
\end{align}

Let $X$ be a standard univariate $t$ distribution.
Then, \eqref{eq:t:E[U|U>z]} implies that
\[\mathbb{E}[X | X > \alpha] =
\left(\frac{\delta}{\delta-2} \right)\left(\frac{t_{\delta-2}\left(\alpha; 0, \frac{\delta}{\delta-2}\right)}{1 - T_\delta\left(\alpha; 0, 1\right)}\right),
\]
for all $\alpha$.
Clearly, $\mathbb{E}[X|X>\alpha] > \alpha$. Hence,
\[\frac{1}{\alpha}\left(\frac{\delta}{\delta-2} \right)\left(\frac{t_{\delta-2}\left(\alpha; 0, \frac{\delta}{\delta-2}\right)}{1 - T_\delta\left(\alpha; 0, 1\right)}\right) > 1.\]
for all $\alpha > 0$.
Setting $(z^* - \mu)/\sigma = \alpha $, which is positive by the definition of $z^*$, in  \eqref{eq:f(z^*)-t} and using the assumption $\mu < 0$,
we conclude that  $f(z^*) > 0$.
\end{proof}

\end{appendix}

\end{document}